\documentclass[twocolumn,showpacs,preprintnumbers,amsmath,amssymb]{revtex4}
\usepackage{epsfig}
\usepackage{dcolumn}
\usepackage{bm}

 \def\bb{\begin{equation}}
 \def\ee{\end{equation}}
 \def\bba{\begin{eqnarray}}
 \def\eea{\end{eqnarray}}
 \def\nn{\nonumber\\}
 
 \let\ds=\displaystyle

 \def\Re{{\rm Re}}

\begin{document}

\title{The evolution of anisotropic structures and turbulence in the
multi-dimensional Burgers equation}

\author{Sergey~N.~Gurbatov}
 \email{gurb@rf.unn.ru}
 \altaffiliation[Also at ]{Observatoire de la C\^ote d'Azur, Lab. Cassiop\'ee,
B.P.~4229, F-06304 Nice Cedex 4, France}
\affiliation{Radiophysics Dept., University of Nizhny Novgorod,\\
23, Gagarin Ave., Nizhny Novgorod 603950, Russia
}

\author{Alexander Yu.~Moshkov}
\affiliation{Radiophysics Dept., University of Nizhny Novgorod,\\
23, Gagarin Ave., Nizhny Novgorod 603950, Russia
}

\author{Alain Noullez}
 \email{anz@obs-nice.fr}
 \affiliation{Observatoire de la C\^ote d'Azur, Lab. Cassiop\'ee,\\
B.P.~4229, F-06304 Nice Cedex 4, France
}

\date{\today}


\begin{abstract}

The goal of the present paper is the investigation of the evolution of
anisotropic regular structures and turbulence at large Reynolds number in the
multi-dimensional Burgers equation. We show that we have local isotropization
of the velocity and potential fields at small scale inside cellular zones. For
periodic waves, we have simple decay inside of a frozen structure.  The global
structure at large times is determined by the initial correlations, and for
short range correlated field, we have isotropization of turbulence. The other
limit we consider is the final behavior of the field, when the processes of
nonlinear and harmonic interactions are frozen, and the evolution of the field
is determined only by the linear dissipation.

\end{abstract}
\maketitle

\section{Introduction}
\label{s:intro}

The well known Burgers equation describes a variety of nonlinear
wave phenomena arising in the theory of wave propagation, acoustics,
plasma physics and so on (see, e.g.,
\cite{Whitham,RudenkoSoluyan,GMS91,WW98,FB2001, BKh2007}). This
equation was originally introduced by J.M.Burgers as a model of
hydrodynamical turbulence \cite{Burgers1939,Burgers1974}. It shares
a number of properties with the Navier--Stokes equation\,: the same
type of nonlinearity, of invariance groups and of energy-dissipation
relation, the existence of a multidimensional version, etc
\cite{Frisch}. However, Burgers equation is known to be integrable
and therefore lacks the property of sensitive dependence on the
initial conditions. Nevertheless, the differences between the
Burgers and Navier-Stokes equations are as interesting as the
similarities \cite{Kr68} and this is also true for the
multi-dimensional Burgers equation\,:
\begin{equation}
 \label{Burg-3D}
 \frac{\partial {\bf v}}{\partial t}  + \left(
 {\bf v}\cdot\nabla \right) {\bf v}  = \nu \nabla^2 {\bf v}\,,
\end{equation}
With  external random forces the multi-dimensional Burgers equation
is widely used as a model of randomly driven Navier-Stokes equation
without pressure
\cite{CheklovYakhot,Polyakov,EKMS,Boldyrev99,DMTRS01}. The
three--dimensional form of  equation  has been used to model the
formation of the large scale structure of the Universe when pressure
is negligible. Known as "adhesion" approximation this equation
describes the nonlinear stage of gravitational instability arising
from random initial perturbation
\cite{GurbatovSaichev1984,GSS89,SZ89,VDFN94,ASG2008}. Other problems
leading to the multi--dimensional Burgers equation, or variants of
it, include surface growth under sputter deposition and flame front
motion \cite{BS95,KM96}. In such instances, the potential ${\psi}$
corresponds to the shape of the front's surface, and the equation
for the velocity potential ${\psi}$ is identical to the KPZ (Kardar,
Parisi, Zhang) equation \cite{BS95,KardarParisiZhang,WW98,BMP95}.
For the deposition problem the velocity in multi--dimensional
Burgers equation ${\bf v}=-\nabla\psi$ is the gradient of the
surface. The mean--square gradient $E(t)=\langle(\nabla {\psi}({\bf
x},t))^2\rangle= \langle {\bf v}^2({\bf x},t)\rangle$ is a measure
of  the roughness of the surface and may either decrease or increase
with time.

When the initial potential $\psi_0({\bf x})$   is a superposition of
one dimensional potentials $\psi_{0,i}(x_i)$, namely
$\psi_0({\bf x})= \sum_i\psi_{0,i}(x_i)$ and $ v_{0,i}({\bf x})=v_{0,i}(x_i)$,
there are no interaction between the velocity component
$v_{0,i}({\bf x},t)=v_{0,i}(x_i,t)$ and  evolution of each component
is determined by one-dimensional Burgers equation.
Before the description evolution of  fields in multi-dimensional Burgers
equation we discuss now very short the evolution of basic types of perturbation
in one-dimensional Burgers equation \cite{Whitham,RudenkoSoluyan,GMS91,WW98},
and compare the behaviour of ``plane'' orthogonal waves in 2-dimensional Burgers equation
with different initial spatial scales.

At infinite Reynolds  number ($\nu\rightarrow0$) the harmonic
perturbation $v_0(x)=k_0\psi_0\sin{k_0x}$ ($\psi_0(x)=\psi_0\cos{k_0x}$),
is transformed at $t\gg t_{\rm nl}=1/k_0^2\psi_0$ into
saw-tooth wave with gradient $\partial v/\partial x=1/t$ and
the same period $L_0=2\pi/k_0$.
It's important that at this stage the amplitude $a=L_0/t$ and the energy
$E(t)=L_0^2/12t^2$
doesn't depend on the initial amplitude. Thus if we compare the
evolution of two components $v_{0,i}(x_i,t)$ with equal potential $\psi_0$
and different scales $L_i$  $(L_1<<L_2)$, the initial energy
will be much higher for the component with smaller
scale $E_1(0)/E_2(0)
=L_2^2/L_1^2$. But asymptotically we have inverse situation $E_1(t)/E_2(t)
\rightarrow{L_1^2}/{L_2^2}$. For large but finite Reynolds number $\Re_0=\psi_0/2\nu$
the shock fronts have a finite width $\sim \nu t/L_0$ and at  $t \gg t_{\rm nl}\Re_0$ we have a
linear stage of evolution where $v(x,t)=4\nu k_0 \sin(k_0x)\exp(-\nu k_0^2t)$.

Continuous random initial fields are also transformed into sequences
of regions (cells) with the same gradient $\partial v/\partial
x=1/t$, but with random position of the shocks separating them. The
merging of the shocks leads to an increase of the integral scale of
turbulence $L(t)$ and because of this the energy $E(t)\sim
L^2(t)/t^2$ of random field decreases more slowly than the energy of
periodic signal. The type of the turbulence evolution is determined
by the behaviour of large scale part of the initial energy spectrum
$E_0(k)\sim \alpha^2k^n$. For $n<1$ the initial potential is
Brownian or fractional Brownian motion and scaling argument may be
used \cite{Burgers1974,GMS91,WW98,VDFN94,SheAurellFrisch,FrMart99}.
In this case the turbulence in self-similar and with integral scale
$L(t)=(\alpha t)^\frac{2}{3+n}$. For $n>1$ the law of energy decay
strongly depends on the statistical properties of the initial field
\cite{WW98,AMS94,EsipovNewman,Esipov,Newman97,Gurbatov2000}.

For an initial Gaussian perturbation the integral scale $L(t)$ and the energy
of the turbulence $E(t)$
\bba
 \label{LtEtD1}
 L(t) &=& (t\sigma_{\psi})^{1/2}\ln^{-1/4}
 \left(\frac{t\sigma_{\psi}}{2\pi l_0^2}\right)\ ,\nn
 E(t) &=& L^2(t)/t^2
\eea are determined only by two integral characteristic of the
initial spectrum\,: the variance of initial potential
${\sigma_\psi}^2=\langle\psi_0^2\rangle$ and the velocity
$\sigma_v^2=\langle v_0^2 \rangle$
\cite{Burgers1974,GMS91,Kida,GurbatovSaichev1981,FournierFrisch,MSW95,GSAFT97,NGAS2005}.
Here $l_0=\sigma_{\psi}/ {\sigma_v}$ is the integral scale of
initial perturbation, and $\sigma_{\psi}/ l_0^2=t_{\rm nl}$ is the
nonlinear time. Thus the energy of two components with equal initial
potential variance $\sigma_{\psi}$ and different scales $l_{0,i},
(l_{0,1}\ll l_{0,2})$ will have very large difference at $t=0$;
${E_1(0)}/{E_2(0)}={l_2^2}/{l_1^2} \gg 1$ and with logarithmic
correction will be the same  at large time ${E_1(t)}/{E_2(t)}\simeq
1$. For large, but finite Reynolds number $\Re_0=\sigma_{\psi}/2\nu$
the shock fronts have a finite width $\sim \nu t/L(t)$  and due to
the multiple merging of shocks the linear regime take place at very
large times $t\gg t_{\rm nl}\exp(\Re_0^2)/\Re_0$. At linear stage
the energy decays as $ Ct^{-3/2}$, where $C \sim L_0
\exp(\Re_0^2)/\Re_0$.

The goal of the present paper is the investigation of the evolution
of anisotropic regular structures and turbulence at large Reynolds
number, when we have a multiple interaction of the spatial
harmonics of the initial perturbation. We shown that we have local
isotropization of the velocity and potential fields inside the
cells. For the periodic wave we have the decay of frozen structure.
The global structure of the random field is determined by the long
correlation of initial field, and for the short correlated field we
have isotropization of turbulence. The other limit we consider in
the paper is the old-age behaviour of the field, when the processes
of nonlinear self-action and harmonic interaction seems to be
frozen, and the evolution of the field is determined only by the
linear dissipation.

The paper is organized as follows. In Section \ref{s:mBurgers} we
formulate our problem and list some results about the solution of
multidimensional Burgers equation in the limit of vanishing
viscosity and its old age behavior. We also shown that we have
local isotropization of the velocity and potential fields. In
Section \ref{s:2D-per} we consider the interaction of plane waves
and evolution of periodic structures in 2-d Burgers equation. In
Section \ref{s:turb} we consider the evolution of anisotropic
multidimensional Burgers turbulence in inviscosid limit. We also
discuss here the influence of finite viscosity and long range
correlation on the late stage evolution of Burgers' turbulence.

\section{Multi-dimensional Burgers equation, the limit of vanishing viscosity
and the old age behaviour}
\label{s:mBurgers}
We shall be concerned with the initial value problem for the
un-forced multi-dimensional Burgers equation (\ref{Burg-3D}) and
consider only the potential solution of this equation, namely
\begin{equation}
\label{potential definition} \bf v (\bf x,t) =
-\nabla\psi(\bf x,t)\ .
\end{equation}
The velocity potential $\psi({\bf x},t)$ satisfies the following nonlinear equation
\begin{equation}
\label{KPZ}
\frac{\partial{\psi}}{\partial t} =
\frac{1}{2}(\nabla{\psi})^2+\nu\nabla^2\psi\ .
\end{equation}
The equation for the velocity potential $\psi$ is identical to the
KPZ (Kardar, Parisi, Zhang) equation
\cite{BS95,KardarParisiZhang,WW98}, which is usually written in the
terms of the variable $h=\lambda^{-1}\cdot\psi$. The parameter
$\lambda$ has the dimension of length divided by time and is the
local velocity of the surface growth. Henceforth $h(\bf x,t)$ has
the dimension of length and is the measure of the surface's
shapeness. In this case ${\bf v}=-\nabla\psi$ is the gradient of
the surface. The roughness of the surface is measured by its
mean-square gradient
\begin{equation}
E(t)=\langle(\nabla {\psi}({\bf x},t))^2\rangle=
\langle {\bf v}^2({\bf x},t)\rangle = \sum_i E_i(t)\ ,
\label{ener}
\end{equation}
Here the angular brackets denote ensemble averages or space
averages for periodic structures. For the one-dimensional
homogeneous field $E(t)$ is the density of energy and always
decreases with time. At the initial stage of evolution and in the
limit of vanishing viscosity the multi-dimensional Burgers equation
is equal to the free motion of the particles. In Lagrangian
representation the velocity of the particle ${\bf V}(t;{\bf y})$ is
a constant. Here ${\bf y}$ is initial (Lagrangian) coordinate of
the particle. In one-dimensional case the increasing of the length
of elementary Eulerian interval $\Delta x = \Delta y + t\Delta V$ is
compensated by the decreasing of length $\Delta x = \Delta y -
t\Delta V$ of the steepening interval and therefore the energy of
the wave (the mean roughness of the curve) is conserved. After
shock formation(s), the energy of the wave decreases with time. In
multi-dimensional case the changing of elementary Eulerian volume
depends also on the initial curvature of perturbation and we don't
generally have compensation of steepening and stretching volumes.
Thus for~$d>1$ the roughness of the surface, measured by its
mean-square gradient $E(t)$ (see~(\ref{ener})) may either decrease or
increase with time \cite{AMS94,Gurbatov2000}.  The increase of the
mean-square gradient in the multi-dimensional Burgers equation (in
contrast with $d=1$) is the result of this equation not having a
conservation form. Nevertheless we will use the expression
"turbulence energy" for the value of $E(t)$ and call
$E_i(t)=\langle(\partial{\psi}/\partial{x_i})^2\rangle =
\langle v_i^2 \rangle$ the energy of the $i$-th velocity component.

Using the Hopf-Cole transformation~\cite{Hopf,Cole}
\begin{equation}
 \label{HC}
 \psi({\bf x},t)=2 \nu \log U({\bf x},t)\ ,
\end{equation}
one can reduce equation~(\ref{Burg-3D}) to the linear diffusion equation
\begin{equation}
 \label{lindif}
 \frac{\partial{U}}{\partial t} = \nu \nabla^2 U\,,\;\;\
 U({\bf x},0) = U_0({\bf x})=\exp\left[\frac{\psi_0({\bf x})}{2\nu}\right]\ .
\end{equation}

The goal of the present paper is the investigation of the evolution of regular
structures and turbulence at large Reynolds number. For the not very large
times we can use the the solutions of Burgers equation in the limit of
vanishing viscosity. The other limit is the old-age behaviour of the field,
when the processes of nonlinear self-action and harmonic interaction seems to
be frozen, end the evolution of the field is determined only by the linear
dissipation. In this case we have the linearisation of Hopf Cole
transformation~(\ref{HC}).

In the limit of vanishing viscosity $\nu\rightarrow0$ use of
Laplace's method leads to the following "maximum representation"
for the potential velocity field \cite{Hopf,GMS91,VDFN94}\,:
\begin{eqnarray}
\psi({\bf x},t)=\max_{{\bf y}}\Phi({\bf x},{\bf y},t)\ ,  \\
\Phi({\bf x},{\bf y},t)=
\psi_0({\bf y})-\frac{({\bf x}-{\bf y})^2}{2t}\ ,
\label{psimax}
\end{eqnarray}
\begin{equation}
{\bf v }({\bf x},t)=\frac{{\bf x}-{\bf y}({\bf x},t)}{t}=
{\bf v }_0({\bf y}({\bf x},t))\ .
\label{vxy}
\end{equation}
Here $\psi_0({\bf y})$ is the initial potential and ${\bf v}_0({\bf
x})=-\nabla\psi_0({\bf x})$. In (\ref{vxy}) ${\bf y}({\bf x},t)$ is
the Lagrangian coordinate where the function $\Phi({\bf x},{\bf
y},t)$ achieves its global or absolute maximum for a given
coordinate ${\bf x}$ and time $t$. It is easy to see that ${\bf y}$
is the Lagrangian coordinate from which starts the fluid particle
which will be at at the point ${\bf x}$ the moment $t$
\cite{GMS91}.

At large times the paraboloid peak in (\ref{psimax}) defines a much
smoother function than the initial potential $\psi_0({\bf y})$.
Consequently, the absolute maximum of $\Phi({\bf x},{\bf y},t)$
coincides with one of the local maxima of $\psi_0({\bf y})$. In the
neighborhood of local maximum $\bf{y_k}$ we can represent the
initial potential in the following form
\begin{equation}
\label{appr}
\psi_0({\bf x})=\psi_{0,k}\left(1-\sum_i\frac{(x_i-y_{i,k})^2}{2L_i^2}\right)\ ,
\end{equation}
where $x_i$ now are the principle axis of the local quadratic form
describing the potential near the local maximum. At relatively
large time the Eulerian velocity field ${\bf v}({\bf x},t)$ in the
whole space will be determined by the particles moving away from
the small regions near the local maximum of $\psi_0(\bf x)$\,:
\begin{equation}
\label{y_i}
{y_i}({\bf x},t)=\frac{(x_i-y_{i,k})}{(1+\psi_{0,k}t/L_i^2)}\ .
\end{equation}
Then, the Lagrangian coordinate ${\bf y}({\bf x},t)$ becomes a
discontinuous function of ${\bf x}$, constant within a cell, but
jumping at the boundaries \cite{GMS91,VDFN94}. The velocity field
${\bf v}({\bf x},t)$ has discontinuities (shocks) and the potential
field $\psi({\bf x},t)$ has gradient discontinuities (cusps) at the
cell boundaries. From (\ref{vxy}),(\ref{y_i}) it becomes clear that
inside the cells the velocity and potential fields have a universal
isotropic and self-similar structure\,:
\begin{equation}
\psi({\bf x},t)=\psi_0({\bf y}_k)-\frac{({\bf x}-{\bf y}_k)^2}{2t}\ .
\label{psimax_k}
\end{equation}
\begin{equation}
{\bf v} ({\bf x},t)=\frac{{\bf x}-{\bf y}_k}{t}\ .
\label{vxy_k}
\end{equation}
One can see that due to the nonlinearity there's the local
isotropisation of the velocity field in the neighborhood of the
local maximum of $\psi_0(\bf x)$. The longitudinal component of the
velocity vector ${\bf v} ({\bf x},t)$ consists of a sequence of
sawtooth pulses, just as in one dimension. The transverse
components consist of sequences of rectangular pulses. At large
times the global structure and evolution of the velocity and
potential fields will be determined by the properties of local
maxima $\psi_0({\bf y}_k)$. For the random field wall motion
results in continuous change of cell shape with cells swallowing
their neighbors and thereby inducing growth of the external scale
$L(t)$ of the Burgers turbulence.

Let us now discuss the old-age limit of the solution of Burgers
equation. Consider a group of perturbation with the bounded initial
potential $\langle \psi_0({\bf x})^2\rangle<\infty$ assuming that
$\psi_0({\bf x})$ is a periodic structure or homogeneous noise with
rather fast decreasing probability distribution of the potential
$\psi_0$. For such a perturbation in $U({\bf x},t)$ we separate a
constant component $\bar{U}$\,:
\begin{equation}
U({\bf x},t)=\bar{U}+\tilde{U}({\bf x},t)
=\bar{U}\left(1+u({\bf x},t)\right) \;\;.
\label{inu}
\end{equation}
Here $u({\bf x},t)=\tilde{U}({\bf x},t)/\bar{U}$ is the relative
perturbation of field $U({\bf x},t)$. Inserting (\ref{inu}) into
(\ref{lindif}) we see that $\bar U$ does not depend on time. Here
$\tilde{U}_0({\bf x})$ and $u_0({\bf x})$ are fields with zero mean
value (on the period or statistically for noise). As times goes on,
the viscous dissipation and oscillation (inhomogeneity) smoothing
causes the amplitude (variance) of the field $\tilde{U}({\bf x},t)$
to become less. At times when its relative amounts $\tilde{U}$ is
small in compare with $\bar{U}$ ($|u|\ll 1$) the solution
(\ref{HC}) is equal to
\begin{equation}
{\bf v} ({\bf x},t)=-2\nu \nabla \tilde{U}({\bf x},t)/\bar
U= -2\nu\nabla u({\bf x},t) \;\;. \label{uhc}
\end{equation}
As $\tilde{U}({\bf x},t)$ and $ u({\bf x},t)$ satisfy the linear
diffusion equation, then $ {\bf v} ({\bf x},t) $ also at these
times fulfills the linear equation. This testifies precisely to the
fact that the evolution of the velocity field enters the linear
stage. The accumulated nonlinear effects are described in this
solution by the nonlinear integral relation between the initial
velocity field $ {\bf v}_0 ({\bf x}) $ and the fields $\tilde U (
{\bf x},0)$, $\bar U$ (\ref{potential definition},\ref{lindif}),
and are characterized by the value $|\Delta\psi_0|/\nu\sim
\Re_0$. Here $\Delta\psi_0$ is the characteristic change in
amplitude of $\psi_0$, and $\Re_0$ is the initial Reynolds number.

From (\ref{uhc}) it is easy to get the well known result, that for
$\Re_0\gg 1$ the initial harmonic waves asymptotically has also
harmonics form, but with the amplitude not depending on the initial
amplitude \cite{Whitham,RudenkoSoluyan}. At large initial Reynolds
number the homogeneous Gaussian field $v_0(x)$ at the nonlinear
stage transforms into series of sawtooth waves with strong
non-Gaussian statistical properties \cite{GMS91,MSW95}.
Nevertheless at very large time, when the relation (\ref{uhc}) is
valid, the distribution of the random field ${\bf v}({\bf x},t)$
with statistically homogeneous initial potential $\psi_0({\bf x})$
converges weakly to the distribution of the homogeneous Gaussian
random field with zero mean value \cite{AMS94} . This stage of
evolution is known as the Gaussian scenario in Burgers turbulence.
In the absence of the long correlation of initial potential field
the potential (and velocity consequently) have an universal
covariance function \cite{GMS91,AMS94}. But the amplitude of this
function is nonlinearly related to the initial covariance function
of the field $\psi_0(\bf x) $ and increases proportionally
$\exp(Re^2_0)$ with increasing of initial Reynolds number $\Re_0$.
When the initial potential has a long correlation ( $\langle (
\psi_0({\bf x})\psi_0(0)\rangle=|{\bf x}|^{-\alpha} F({\bf x}/x)$
$, 0<\alpha<3 $ ) at large $x$ we have conservation at linear stage
as long correlation as the anisotropy of the field $F({\bf x}/x)$
\cite{AMS94} .

\section{The evolution of periodic
structures  and the interaction of plane waves in 2-d
Burgers equation } \label{s:2D-per}

It was shown in previous section that at large time we have a
cellular structure of the field with universal behaviour of
potential and velocity inside each cell. The global structure of
the field will be determined by the properties of local maximum of
initial potential.

Let us consider the evolution of periodic structure in 2-dimensional Burgers equation
\begin{equation}
\label{perstr}
\psi_{0}(x_1,x_2)=2\psi_0\cos(k_2x_2)\cos(k_1x_1).
\end{equation}
The evolution of this structure my be interpreted as the
interaction of two plane waves with equal modulus of the wave number
$k_0$
\begin{equation}
\label{twoplane}
\psi_{0}(x_1,x_2)=\psi_0\cos(k_1x_1+k_2x_2)+\psi_0\cos(k_1x_1-k_2x_2).
\end{equation}
Here $k_1=k_0n_1=2\pi/l_1$, $k_2=k_0n_2=2\pi/l_2$ and $\bf n $ is
unit vector with the component $n_1,n_2$. For the each of
noninteracting plane waves the energy of $i$-th component is
$E_{i,p}(t)=E(t)n_i^2$, where $E(t)$ is the energy of plane wave
and $E(0)=\psi_0^2k_0^2/2$. For $\nu\rightarrow0$ the plane wave is
transformed at $t\gg t_{\rm nl}=1/k_0^2\psi_0$ into saw-tooth wave with
gradient $\partial v/\partial x=1/t$ and the energy
$E(t)=\pi^2/3k_0^2t^2$. For large but finite Reynolds number we
have a linear stage of evolution where $v(x,t)=4\nu k_0
\sin(k_0x)\exp(-\nu k_0^2t)$.

Consider first the late stage of evolution of periodic structure of
times $t\gg t_{\rm nl}$. At this stage the velocity has the
universal form in each cell (\ref{vxy_k}), where ${\bf y_k}$ are the
maximums of the initial potential (\ref{perstr}). For the initial
potential there are two sets of maximum of equal value corresponding
the conditions $\cos{k_1x_1}=\cos{k_2x_2}=1$ and
$\cos{k_1x_1}=\cos{k_2x_2}=-1$. The shock lines (cell boundaries) of
the velocity field are orthogonal to vector connecting the neighbor
cell center and they are immobile and situated in the middle between
the centers. Due to the symmetry of initial conditions the velocity
field is symmetric over the point $(l_1/2,l_2/2)$. Assume now that
$l_1\leq l_2$ and consider the velocity fields inside the region
$S$\,: $\left(x_1\in [0,l_1/2], x_2\in [0,l_2/2] \right)$, see
figure \ref{fig1}.

\begin{figure}
\epsfig{figure=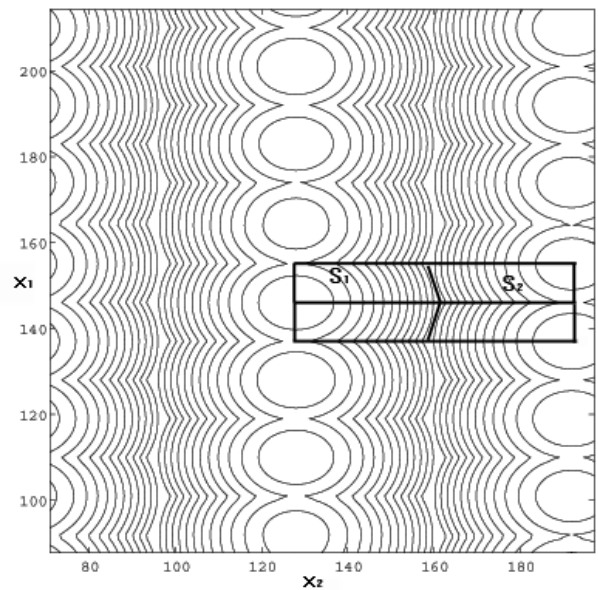, width=8truecm} \caption{Cellular structure
of potential for $k_1 \gg k_2$.} \label{fig2}
\end{figure}

The region $S$ is divided by the shock line in the regions $S_1$
and $S_2$
\begin{equation}
\label{S_1}
S_1:\left(0\leq x_2\leq -\frac{l_1}{l_2}x_1+
\frac{l_1^2+l_2^2}{4l_2^2}\right)\;\;.
\end{equation}
The center of the cell inside the region $S_1$ is in the point
$(x_1=0, x_2=0)$ and inside the region $S_2$ is in the point
$(x_1=l_1/2,x_2=l_2/2)$. Consequently for the velocity fields one
can get
\begin{eqnarray}
v_1=x_1/t,\;\;\ v_2=x_2/t, \;\;\; \;{\bf x} \in S_1,\;\;
\nonumber \\ v_1=(x_1-l_1/2)/t,\;\;\ v_2=(x_2-l_2/2)/t,\;\;\;{\bf x} \in S_2.
\label{vel}
\end{eqnarray}
Thus on large time we have a frozen structure of the field with decreasing
amplitude $ \sim  t^{-1}$. For the energy of the velocity component from
expressions (\ref{vel}) one can obtain
\begin{eqnarray}
E_1(t)=\frac{l_1^2}{12t^2}\left(1-\frac{l_1^2}{l_2^2}\right)=
\frac{\pi^2}{3k_0^2n_1^2t^2}\left(1-\frac{n_2^2}{n_1^2}\right),
\nonumber
\\E_2(t)=\frac{l_2^2}{48t^2}\left(1+\frac{l_1^4}{l_2^4}\right)=
\frac{\pi^2}{12k_0^2n_2^2t^2}\left(1+\frac{n_2^4}{n_2^1}\right).
\label{per-en}
\end{eqnarray}

From (\ref{S_1}),(\ref{vel}) one can receive that for the very
anisotropic fields $(l_1 \ll l_2), (n_1 \simeq 1,n_2 \ll 1) $ the
velocity component $v_1$ reproduce the behaviour of the velocity in
one-dimensional Burgers equation, but the large scale component has
now the period $L=l_2/2$ instead of $L=l_2$ for the initial
perturbation. Let us compare the decay of the periodic structure
(\ref{perstr}), which is a superposition of two plane waves with
the decay of the energy component $E_{i,p}$ of singular plane wave.
For $l_1 \ll l_2$ the energy of small scale decays as $E_1\simeq
\pi^2/3k_0^2n_1^2t^2 \simeq E_{i,1}=\pi^2n_1^2/3k_0^2 t^2$. The
energy of large scale component $E_2\simeq \pi^2/12k_0^2n_2^2t^2
\gg E_{i,2}=\pi^2n_2^2/3k_0^2t^2$.

Consider now the linear stage of the evolution of the periodic
structure, when the Hopf-Cole solution is reduced to the linear
relation (\ref{uhc}) between the velocity field ${\bf v({\bf
x},t)}$ and the solution $\tilde{U}({\bf x},t)$, $u({\bf x},t)$ of
the linear diffusion equation (\ref{lindif}). Using the relation $
\exp{(z\cos{\theta})}=I_0(z)+2\sum_{m=1}^\infty I_m(z)\cos{m\theta}$,
where $I_m(z)$ are a modified Bessel functions, for the solution of
this equation we have from (\ref{lindif}),(\ref{perstr})
\begin{eqnarray}
\label{solperU}
U({\bf x},t) = &\ds I_0^2(\Re_0)\hfill\nonumber \\
&\ds +2\sum_{m=1}^\infty & I_0(\Re_0) I_m(\Re_0)\hfill\nonumber \\
&&\{\cos[m(k_1x_1+k_2x_2)]\nonumber \\
&&+\cos[m(k_1x_1-k_2x_2)]\}\nonumber \\
&&{\rm e}^{-\nu m^2(k_1^2+k_2^2)t}\nonumber \\
&\ds +2\sum_{n=1}^\infty \sum_{m=1}^\infty &I_m(\Re_0) I_n(\Re_0)\hfill
\nonumber \\
&&\{\cos[(n+m)k_1x_1+(n-m)k_2x_2]\nonumber \\
&&{\rm e}^{-\nu ((n+m)^2k_1^2+(n-m)^2k_2^2))t}
\nonumber \\
&&+\cos[(n-m)k_1x_1+(n+m)k_2x_2)]\nonumber \\
&&{\rm e}^{-\nu ((n-m)^2k_1^2+(n+m)^2k_2^2)t}\}\ ,
\end{eqnarray}
where $\Re_0=\psi_0/2\nu$. Here the first sum described the
nonlinear evolution of two plane waves and the double summation -
the interaction between the plane waves. From eq.(\ref{solperU}) we
have that a constant component in eq. (\ref{inu}) is
$\bar{U}=I_0^2(\Re_0)$ and $\tilde{U}({\bf x},t)= U({\bf
x},t)-\bar{U}$. At linear stage of evolution, when $u({\bf
x},t)=\tilde{U}({\bf x},t)/\bar{U}\ll 1$ we have from eq.
(\ref{HC}),(\ref{inu})
\begin{equation}
\label{ulindef}
\psi({\bf x},t)=\bar{\psi}+ \tilde {\psi}({\bf x},t),\,\,\tilde {\psi}({\bf x},t)\simeq 2\nu u({\bf x},t),
\end{equation}
where $\bar{\psi}= 2\nu \log\bar{U}$. The asymptotic behaviour of
the potential (shape of the surface) $\tilde {\psi}({\bf x}) $ will
be determined by the low index of the decaying exponent in solution
(\ref{solperU}). For $3^{1/2}k_2>k_1>k_2$ we have
\bba
\label{ulin}
\tilde {\psi}({\bf x},t) \simeq&&
4\nu (I_1(\Re_0)/I_0(\Re_0))\nn
&\times&\cos{(k_1x_1)}\cos{(k_2x_2)}e^{-\nu(k_1^2+k_2^2)t}  \:\;\,
\eea
and consequently the velocity component $v_i$ decays like the $i$-th component
of the velocity of plane wave
\bba
\label{v1lin}
v_1({\bf x},t)=&&
4\nu k_1 (I_1(\Re_0)/I_0(\Re_0))\nn
&\times&\sin{(k_1x_1)}\cos{(k_2x_2)}e^{-\nu(k_1^2+k_2^2)t}  \:\;\,
\eea
\bba
\label{v2lin1}
v_2({\bf x},t)=&&
4\nu k_2 (I_1(\Re_0)/I_0(\Re_0))\nn
&\times&\sin{(k_2x_2)}\cos{(k_1x_1)}e^{-\nu(k_1^2+k_2^2)t}  \:\;\,
\eea
For the small Reynolds number this solution is equal to the linear
solution of Burgers equation. But for $3^{1/2}k_2<k_1$ the
nonlinear interaction between the plane waves change the asymptotic
evolution of the potential (shape of the surface). Now the leading
term in eq. (\ref{solperU}) is along axis $x_1$
\bba
\label{ulin2}
\tilde {\psi}({\bf x},t) \simeq&&
2\nu  (I_2(\Re_0)/I_0(\Re_0))\nn
&\times&\cos{(2k_2x_2)}e^{-\nu(4k_2^2)t}  \:\;\,
\eea
and is on the second harmonic along axis $x_2$. Thus
\bba
\label{v2lin2}
v_2({\bf x},t)=&&
4\nu k_2 (I_2(\Re_0)/I_0(\Re_0))\nn
&\times&\sin{(2k_2x_2)}e^{-\nu(4k_2^2)t}  \:\;\,
\eea
and we have depressing of the modulation of potential along axis
$x_1$. The evolution of the gradient of surface along $x_1$ (velocity
component $v_1$ is still determined by eq. (\ref{v1lin}) and they
decay faster than the gradient over $x_1$.

Thus due to the nonlinearity which leads to the generation of
cross-wave numbers we have for the velocity component $v_2$ at
linear stage instead of initial spatial frequency $k_2$ the leading
term at the second harmonic. This one is true even for the small
Reynolds number. For the large Reynolds number $\Re_0$ we have
$I_1(\Re_0)/I_0(\Re_0)\simeq 1$ and the amplitude of $v_1$,$v_2$ do
not depend on the amplitude of initial perturbation.

Let us consider the transition processes of very anisotropic field
when the angle between interacting plane waves is small $n_1 \gg
n_2 $. Consider first a more general situation when plane periodic
is modulated by large scale function $M({\bf x})$ and the initial
potential is represented as
\begin{equation}
\psi_0^M({\bf x})=M({\bf x})\psi_0\cos{(k_1x_1}) \;\;.
\label{psiMFcos}
\end{equation}
We assume that the function $M({\bf x})$
characterized by the scales $L_{M,i}$ and $L_{M,i}\gg l_1=1/2\pi k_1$, For the
plane interacting waves (\ref{perstr})
$M({\bf x})=2\cos(k_2x_2)$. For the initial velocity field we
have from  eq. (\ref{psiMFcos})
\begin{eqnarray}
v_{1,0}(x_1,x_2)\simeq k_1\psi_0\sin(k_1x_1)M(x_1,x_2),
\\ v_{2,0}(x_1,x_2)=-\psi_0\cos(k_1x_1)M^{'}_{x_2}(x_1,x_2).
\label{v-per}
\end{eqnarray}
In the limit of vanishing viscosity $\nu\rightarrow0$ the evolution
of the velocity field is described by the equation (\ref{vxy}) and
${\bf v }({\bf x},t)= {\bf v }_0({\bf y}({\bf x},t))$, where ${\bf
y}({\bf x},t)$ is Lagrangian coordinate from which starts the fluid
particle which will be at at the point ${\bf x}$ the moment $t$.
While $L_{M,i}\gg l_1$ the velocity component $v_2\ll v_1$ and at
$t\ll t_{\rm nl,2}=L_{m,2}^2/\psi_0$ one can neglect the drift of the
particles along $x_2$ axis. In this case for the Lagrangian
coordinates we get
\bba
X_1(y_1,y_2,t)=y_1+tk_1\psi_0\sin(k_1y_1)M(y_1,x_2)\ ,\nn
X_2(y_1,y_2,t)=y_2\ .
\label{lagr-x-init}
\eea
Before $t<t_{{\rm nl},1}=1/k_1^2\psi_0$ this solution ${\bf v }({\bf
x},t)= {\bf v }_0({\bf y}({\bf x},t))$ is single-valued. For
$t>t_{{\rm nl},1}$ we need to introduce the shock in multi-valued
solution. In quasi--static approximation we assume that the
evolution of the velocity component $v_1(x_1,x_2,t)$ is equal to
the evolution of initial harmonic perturbation
$v_{1,0}=A\sin(k_1x_1)$ in one-dimensional Burgers equation. The
amplitude of the perturbation $A=k_1\psi_0 M(x_{1,m},x_2)$
($x_{1,m}=l_1m$) depends on the coordinate $x_2$ as a parameter and
is assuming to be a constant of $x_1$ on the each period of the
harmonic perturbation.

Let's now consider the nonlinear stage of evolution $t_{{\rm nl},1} \ll t
\ll t_{{\rm nl},2}$ when the velocity component $v_1$ transforms into
sawtooth waves. Consider the region when $M(x_1,x_2)>0$. It's easy
to see that at $t\gg t_{{\rm nl},1}$ each period
$l_1(m-1/2)<x_1<l_1(m+1/2) $ will be cover by the particles from
small region near the point $x_{1,m}=l_1m$ and the solution of the
equations $x_1=X(y_1,y_2,t)$, $x_2=X_2$ may be written as
\begin{equation}
\label{y1}
y_1-x_{1,m}=\frac{x_1-x_{1,m}}{1+tk_1^2\psi_0M(x_{1,m},x_2)},\;\;y_2=x_2.
\end{equation}
The shocks are  situated at the line  $x_{1,s}=l_1(m+1/2)$. From (\ref{lagr-x-init}- \ref{y1})
one can receive for the velocity
component at $l_1(m-1/2)<x_1<l_1(m+1/2)$  and for $M>0 $
\bba
v_1(x_1,x_2,t)&=&\frac{x_1-y_{1,m}}{t}\nn
&&\left(1-\frac{1}{1+tk_1^2\psi_0 M(x_{1,m},x_2)}\right)\ ,\nn
v_2(x_1,x_2,t)&=&-\psi_0M^{'}_{x_2}(x_{1,m},x_2)\nn
&&\cos\left(\frac{k_1x_1}{tk_1^2\psi_0M(x_{1,m},x_2)}\right)\ .
\label{v_i-per}
\end{eqnarray}
From this equation one can see that for $t\gg t_{{\rm nl},1}$ the
velocity component $v_1(x_1,x_2,t)$ is transformed into saw-tooth
wave $v_1\simeq (x_1- y_{1,m})/t$ like in one-dimensional case. It
means that we have a fully depression of initial amplitude
modulation of this component. The velocity component $v_2$ loss the
periodic modulation over $x_1$ and
$v_{2,0}(x_1,x_2,t)\simeq-\psi_0M^{'}_{x_2}(x_1,x_2)$ for positive
$M$ and $v_{2,0}(x_1,x_2,t)\simeq\psi_0M^{'}_{x_2}(x_1,x_2)$ for
negative $M$. The energy of this component increases twice in
compare with the initial energy.

For such wave  the velocity field  may be also represented  in the form
\begin {equation}
\label{veltwo}
{\bf v} ({\bf x},t) ={\bf v}_{l} ({\bf x},t)+{\bf v}_s ({\bf x},t) ,\;\;\;
{\bf v}_{l} ({\bf x},t) =\langle {\bf v} ({\bf x},t)\rangle
\end{equation}
where the brackets $\langle ...\rangle$ means the averaging over
period $l_1$ . Here ${\bf v}_{l} ({\bf x},t)$ is the large scale
component and ${\bf v}_s ({\bf x},t)$ is the small scale component.
We assume that the evolution of small scale component ${\bf v}_s$
may be described in the quasi-static approximation. The mean
velocity ${\bf v}_{l}$ has the scale in order of $L_{M,i}$, and at
stage $t \ll t_{{\rm nl},M}= L_{M,i}^2/ \sigma_{\psi}$ one can neglect
nonlinear distortion and dissipation of this component. Then from
eq. (\ref{KPZ}) we have
\begin{equation}
\label{velmean}
\frac{\partial {\bf v}_{l}({\bf x},t)}{\partial t} =
  -\frac{1}{2}\nabla\langle  {\bf v}_s^2 ({\bf x},t) \rangle =
 -\frac{1}{2}\nabla E_s({\bf x},t) \;\ ,
\end{equation}
The integration of this equation over $t$ give the evident
expression for the coherent component
\begin{equation}
\label{velmeanpsi}
{\bf v}_{l} ({\bf x},t)=-\nabla\langle\psi ({\bf x},t) \rangle
\end{equation}
Here we have used the equation (\ref{KPZ}) for the potential $\psi
({\bf x},t)$. From eq. (\ref{velmean}) we have that the generation
of large scale component is determined by the gradient of the
energy of small scale component. For the periodic modulation
$E_s(x,t)$ at $t\gg t_{{\rm nl},1}$ does not depend on the initial
amplitude. It means that at this times there are no generation of
the large scale component. The gradient of mean potential
$\langle\psi ({\bf x},t)\rangle=\psi_0 |M({\bf x})|-l_1^2/16t$ at
this time does not depend on $t$ and from eq. (\ref{velmeanpsi}) we
get
\begin{equation}
\label{velmeanpsiper}
{\bf v}_{l} ({\bf x},t)=-\nabla\langle\psi ({\bf x},t) \rangle=-\psi_0\nabla |M({\bf x})|.
\end{equation}
The amplitude of the small scale component is $l_1/t$, while the
amplitude of the large scale component is in order $\psi_0/L_{M,i}$
and it means that at $t>L_{M,i}l_1/\psi_0$ the main part of energy
is in the large scale component. The nonlinear distortion of large
scale component is significant at $t>min(L_{M,i}^2)/\psi_0$. The
future evolution of this component is strongly depend of the
properties of modulation function $M({\bf x})$.

\begin{figure}
\begin{minipage}[c]{.45\linewidth}
\epsfig{figure=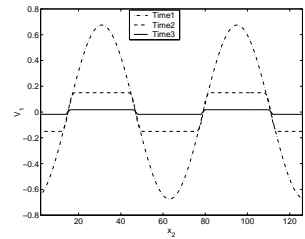, width=\linewidth}
\caption{Evolution of $v_1(x_2,t)$.} \label{v1x2}
\end{minipage}
\end{figure}

\begin{figure}
\begin{minipage}[d]{.45\linewidth}
\epsfig{figure=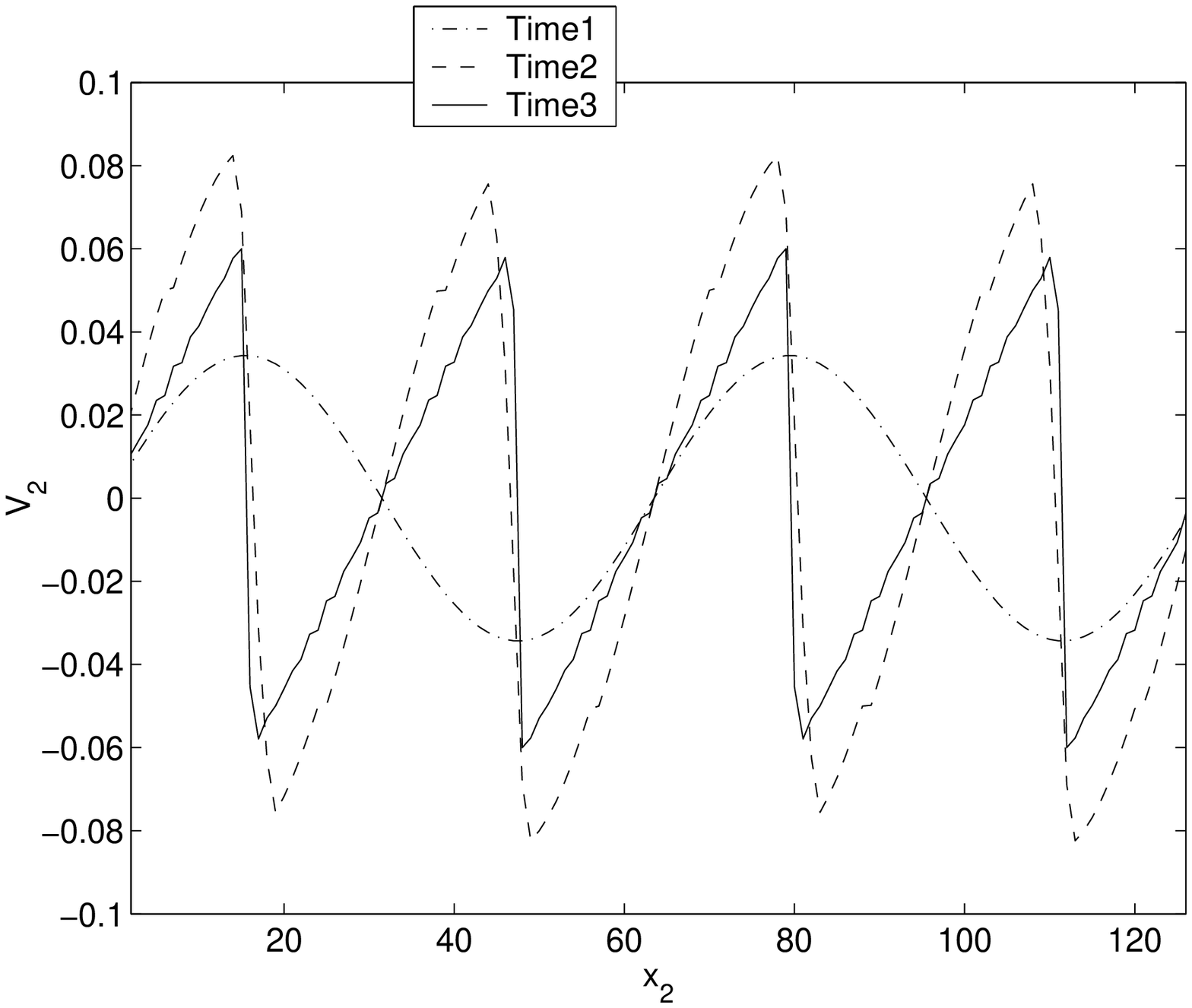, width=\linewidth}
\caption{Evolution of $v_2(x_2,t)$.} \label{v2x2}
\end{minipage}
\end{figure}

\begin{figure}
\begin{center}
\epsfig{file=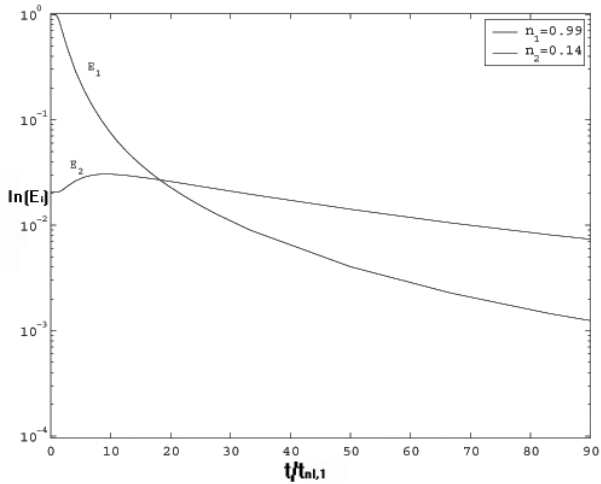,width=8truecm}
\caption{Evolution of the velocity components energies.}
\label{fig5}
\end{center}
\end{figure}

For the periodic structure (\ref{perstr}) when $k_1 \gg k_2$ we
have that the plane wave $\psi_0\cos(k_1x_1)$ is modulated by large
scale function $M({\bf x})=\cos{(k_1x_1)}$ eq. (\ref{psiMFcos}).
The initial perturbation in this case is periodic structure with
periods $l_1=2\pi/k_1$ , $l_2=2\pi/k_2$ and $l_1 \ll l_2$.

Before the nonlinear distortion of large scale component $t \ll
t_{{\rm nl},2}=l_2/\sigma_{\psi}$ the evolution of structure take
place like in general case of modulated wave eq. (\ref{psiMFcos}).
The velocity component $v_1(x_1,x_2,t)$ is transformed into
saw-tooth wave (Figure \ref{v1x2})  and we have a fully depression
of initial amplitude modulation of this component (Figure
\ref{v1x2}). The velocity component $v_2$ loss the periodic
modulation over $x_1$ and
$v_{2,0}(x_1,x_2,t)\simeq\psi_0|\cos{(k_2x_2)}|{'}_{x_2}$. The
period of this component is twice less than the initial period
(Figure \ref{v2x2}) and the energy increases twice in compare with
the initial energy. The behavior of the energy is shown on the
Figure \ref{fig5}.


\section{Evolution of anisotropic  multidimensional Burgers turbulence}
\label{s:turb}
\subsection{The intermediate stage of evolution of anisotropic random field}
\label{s:turbin}

In this section we will consider the intermediate stage of
evolution of anisotropic random field in the two dimensional
Burgers equation. Let us assume that the initial potential
$\psi_0(x_1,x_2)$ is random and strong anisotropic field with the
spatial scales $l_1 \ll l_2$. The initial energy $E_i(t)=\langle
v_i^2\rangle=
\langle(\partial\psi/\partial{x_i})^2\rangle = \sigma_{\psi}/l_i^2$ of
the velocity component $v_1$ is in this case much larger than the
energy of the large scale component $v_2$ . We can introduce the
nonlinear time of $i$-th component as
$t_{{\rm nl},i}=l_i^2/\sigma_{\psi}$. For $t \ll t_{{\rm nl},2}$, the drift of
the Lagrangian particles in direction $x_2$ is relatively small.
Then one can assume $y_2=x_2$ in eq. ({\ref{psimax}}) and consider
the one-dimensional problem with the initial potential
$\psi_0(y_1,x_2,t)$, where $x_2$ is a parameter.

\begin{figure}
\centering \epsfig{figure=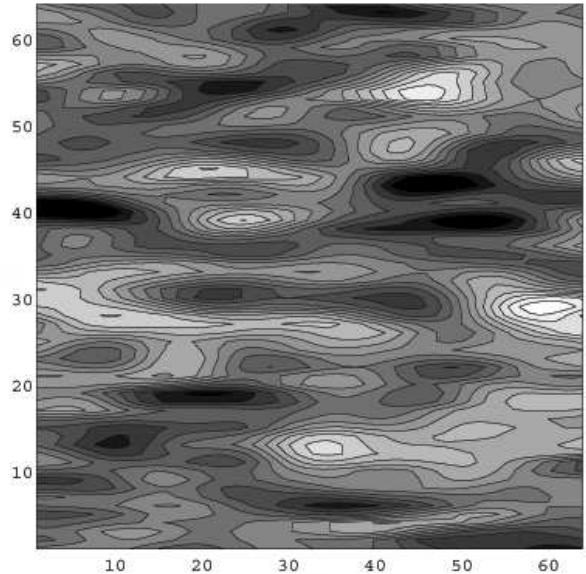, width=8truecm}\\
\centering \epsfig{figure=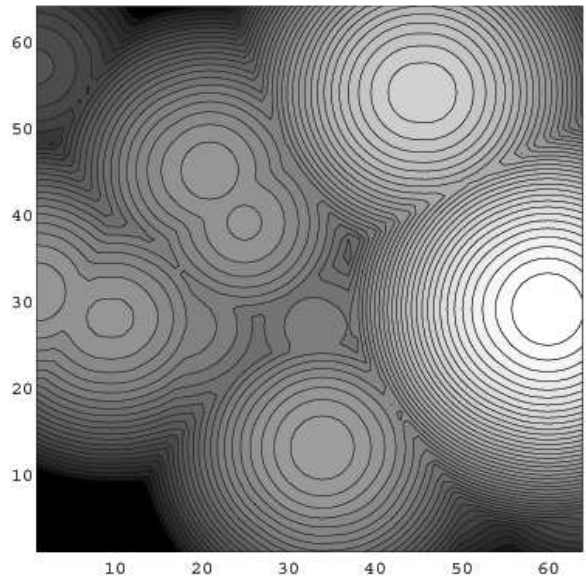, width=8truecm}
\caption{Evolution  of potential. (Initial potential and stage of
isotropization) } \label{fig6}
\end{figure}

Due to the condition $l_1 \ll l_2$ the the first shock lines in the
Lagrangian coordinates are on the points where
$\partial{\psi}/\partial{y_1}$ has a minimum. In Eulerian space the
are oriented primarily along the $x_2$ axis end its length in this
direction increase in time. For $t\gg t_{{\rm nl},1}$ the velocity field
$v_1(x_1,x_2,t)$ transforms to the sequence of triangular pulses
\bba
\label{triangle}
v_1(x_1,x_2,t)&=&\frac{x_1-y_{1,k}(x_1,x_2,t)}{t}\ ,\\
&&{\rm for}\quad x_{1,k}^s<x_1<x_{1,k+1}^s\ ,\nonumber
\eea
where $y_{1,k}(x_1,x_2,t)$ are the coordinates of absolute maximum
of (\ref{psimax}) over $y_1$ with $y_2=x_2$. The shock positions
$x_{1,k}^s(x_2,t)$ are
\bba
\label{shock}
x_{1,k}^s&=&\frac{y_{1,k+1}+y_{1,k}}{2}+v_k t\ ,\\
v_k&=&\frac{\psi_0(y_{1,k}(x_1,x_2),x_2)-\psi_0(y_{1,k+1}(x_1,x_2),x_2)}
{y_{1,k+1}(x_1,x_2)-y_{1,k}(x_1,x_2)}\ .\nonumber
\eea
It means that at fixed $x_2$ the interval
$x_{1,k}^s<x_1<x_{1,k+1}^s$ will be cover by the particles from
small region near the Lagrangian point $y_{1,k}(x_1,x_2,t)$, and
for the velocity component $v_2$ we get
\bba
\label{v2}
v_2(x_1,x_2,t)&=&
-\frac{\partial{\psi_0}(x_1,x_2)}{\partial{x_2}}|_{x_1=y_{1,k}(x_2,t)}\ ,\\
&&{\rm for}\quad x_{1,k}^s<x<x_{1,k+1}^s\ .\nonumber \eea The
velocity $v_2(x_1,x_2,t)$ doesn't depend on $x_1$ between the
shock-lines $x_{1,k}^s(x_2)$ and $x_{1,k+1}^s(x_2)$. The collision
of the shocks in one-dimensional Burgers equation is now equal that
at some point $ {\bf x}_{*} $ two adjacent shock lines $x_{1,k}^s$
and $x_{1,k+1}^s$ touch each other. Then this point will be
developed into new shock lines $x_{1,k}^{s,*}(x_2)$ with the
increasing in time length along axis $x_2$ and which on its ends
transforms into lines $x_{1,k}^s$,$x_{1,k+1}^s$. Thus at
$t_{n2,1}\gg t\gg t_{{\rm nl},1}$ the velocity field has a cellular
structure, the border $x_{1,k}^s(x_2)$ of the cells are describing
by the equation (\ref{shock}), the velocity component $v_1({\bf x})$
has an universal structure (\ref{triangle}). Velocity component
$v_2({\bf x})$ inside the cell does not depend on $x_1$ and along
the axis $x_2$ reproduced the behaviour of $v_2$ along the line
$x_1=y_{1,k}(x_2,t)$ eq. (\ref{v2}). The evolution of the potential
is plotted on Figure \ref{fig6}.

The statistical problems of the velocity component $v_1$ at this
stage are similar to the properties of one-dimensional Burgers
turbulence. The integral scale $L_1(t)$ and the energy $E_1(t) \sim
\sigma_{\psi}/t$ of the component $v_1$ are described by the
expressions (\ref{LtEtD1}), where $\sigma_{\psi}$ is the variance of
initial two-dimensional potential $\psi_0$ and $l_0=l_1$ is the
integral scale of $v_1$ component $L_1=\sigma_{\psi}/\sigma_{v_1}$.
Due to the merging of the shock lines the integral scale of the
turbulence along the axis $x_1$ increases with time $L_1(t)\sim
(t\sigma_{\psi})^{1/2}$ and at $t \sim t_{{\rm nl},2}$, when
$L_1(t)\simeq l_2 $ and $E_1(t)\simeq E_2(0)$ we need to take into
account the nonlinear distortion along the axis $x_2$. At $t \gg
t_{{\rm nl},2}$ the potential and velocity fields have a universal
isotropic and self-similar structure inside the cells:
eq.(\ref{psimax_k}), (\ref{vxy_k}). The boundary of the cells on
this stage degenerate into straight lines (planes, in three
dimensional case). The multiply merging of the cell will leads to
the establishment of statistical self-similarity and isotropization
of the field. In the next section we will show how the statistical
properties of the isotropic multidimensional Burgers turbulence are
connected with the parameter of anisotropic initial perturbation.

\subsection{ Isotropisation of the multidimensional Burgers turbulence}
\label{s:isotrop}

The statistical properties of the Burgers velocity field $\bf v$
(equation (\ref{vxy})) in the limit $\nu\rightarrow 0$ are
determined by the statistical properties of the absolute maximums
coordinate $\bf {y}(\bf {x},t)$ of the function (\ref{psimax}). In
one-dimensional case the problem of the absolute maximum is reduced
to the problem of the crossing the random signal $\psi_0(x)$ by the
non homogeneous function $(x-y)^2/2t+H$. The asymptotic behaviour
of the field at large $t$ is determined by the maximum which
amplitude is higher than the variance of the initial potential.
That's why one can use some results of the theory of extremal
processes \cite{LLR83,GMS91,MSW95}. In multidemensional case the
problem of the peaks statistic of the Gaussian field is rather well
known for the isotropic and homogeneous field \cite{BBKS86}. But
for the Burgers turbulence we need to find the statistical
properties of the absolute maximum of the scalar non homogeneous
and anisotropic field $\Phi({\bf x},{\bf y}, t)$. In paper
\cite{GM2003} it was shown that at large $t$ this problem is
reduced to the problem of the finding of the statistical properties
of extremum of random field $\psi_0({\bf x})$ which value is much
greater than their variance $\sigma_{\psi}$.

Let us assume the initial potential $\psi_0({\bf x})$ is random
Gaussian field and it's correlation function may be used in the
following form
\begin{equation}
\label{corr}
\langle\psi_0({\bf x})\psi_0({\bf x}+{\bf z})\rangle =B_{\psi}({\bf z})=\sigma_{\psi}^2
\prod_{i=1}^d{R_i(z_i)},
\end{equation}
\begin{equation}
\label{R-ap}
R_i(z_i)=1-\frac{z_i^2}{2!l_{0,i}^2}+\frac{z_i^4}{4!l_{1,i}^4}+...  .
\end{equation}
We assume also that the correlation function decreases rather fast
at large distances; $B_{\psi}(|{\bf z}|>l_{st})\simeq 0$. Then the
Gaussian initial field $\psi_0({\bf x})$ in the points $|{\bf
x_1}-{\bf x_2}|>l_{st}$ is statistically independent.

In the limit of vanishing viscosity we have ''maximum
representation'' for the potential eq. (\ref {psimax}). In this
solution the velocity field ${\bf v}({\bf x},t)$ eq. (\ref{vxy}) is
determined by the coordinate $\bf y$ of the absolute maximum of
function $\Phi({\bf x,y},t)$. Let $Q(H,\triangle V_k)$ denote the
cumulative probability and $W_{max}(H,\triangle V_k)$ denote the
probability density of the absolute maximum in elementary volume
$\triangle V_k$
\begin{equation}
\label{Q}
Q(H,\triangle V_k)=Prob(\Phi <H, {\bf y}\in\triangle V_k),
\end{equation}
\begin{equation}
\label{dens}
W_{max}(H,\triangle V_k)=Q'_H(H,\triangle V_k).
\end{equation}
Here we introduce the elementary volume  $\triangle V_k$ which scale is much greater
than $l_{st}$, but much smaller than the integral scale of turbulence $L(t)$.
The probability for the absolute maximum to be contained between $H_1$
and $H_1+\triangle H_1$ with the coordinate ${\bf y}({\bf x},t)\in\triangle V_k$ equals to
that one for the absolute maximum $H_1\in (H,H+\triangle H)$ to lie in
$\triangle V_k$ and to be less than $H$ in magnitude for outer
intervals $\overline{\triangle V}_k$
\bba
\label{prob}
&Prob&\ds({\bf y}\in\triangle V_k, H\in[H_1,H_1+\triangle H])\nn
&&=W_{max}(H,\triangle v_k)\triangle H Q(H,\overline{\triangle V}_k).
\eea
Here we propose that the absolute maximums are statistically
independent in the intervals $\triangle V_k$ and
$\overline{\triangle V}_k$ . The probability for coordinate ${\bf
y}({\bf x},t)$ to fall into $\triangle V_k$ can be obtained by the
integration of (\ref{prob}) with respect of H
\begin{equation}
\label{ind-prob}
Prob({\bf y}\in\triangle V_k)=\int{W_{max}(H,\triangle V_k)Q(H,\overline{\triangle V}_k)}dH.
\end{equation}
After the integration of (\ref{ind-prob}) by
parts we have
\begin{equation}
\label{fin-prob}
Prob({\bf y}\in\triangle V_k)=\int{N(H,\triangle V_k)Q'_H(H)}dH,
\end{equation}
where Q(H) is the integral distribution function of absolute
maximum in the whole space. In Appendix in \cite{GM2003} it was
shown that for large $H$ the integral distribution function
\begin{equation}
\label{probdis}
Q(H,\triangle V_k)=\exp{\left(-N(H,\triangle V_k)\right)}
\end{equation}
is determined by the mean number of
extremum $N(H,\triangle V_k)$ with value larger than $H$.

Let us first consider the statistical properties of the extremum of
the homogeneous random field $\psi_0({\bf x})$. It's known that for
the smooth fields the number of crossing some higher level
asymptotically tends to the number of maximum and number of
extremum. It means, that all the peaks above some high lever have
only one extremum, which is the maximum of this peak. Thus we will
consider first the properties of extremum of the field $\psi_0({\bf
x})$. Using these properties of $\delta$-function one can obtain
for the mean number $ N(H;V)=\langle N_{exp} \rangle$ of extremum
with the value greater than $H$ in some volume $V$
\begin{equation}
\label{grval}
N(H;V)=\left\langle \int_V \delta(\nabla \psi_0) |J(a_{ij})|E(\psi_0-H)d{\bf x}
\right\rangle.
\end{equation}
Here $E(s)$ is a unit function, and $J$ is the Jacobian of
transformation
\begin{equation}
\label{Jacob}
J(a_{ij})=|a_{ij}|,\;\;\;
a_{ij}=\frac{\partial{\psi_0({\bf x})}}{\partial{x_i}\partial{x_j}}.
\end{equation}
For the homogeneous field $\psi_0({\bf y})$ one can introduce the
density of extremum as $n_{ext}(H;V)=N(H)/V$. The density of
$n_{ext}$ in this case is determined by the joint probability
distribution function of the $\psi_0$, their gradient $v_i=\partial
{}\psi_0/\partial {x_i}$ and tensor $a_{ij}=\partial
{\psi_0}/\partial{x_i}\partial{x_j}$. For the homogeneous Gaussian
field
$W_{\psi_0,v_i,a_{ij}}=W_v(v_i)W_{\psi_0,a_{ij}}(\psi_0,a_{ij})$
and from eq. (\ref{grval}) we have for the density of extremum
\begin{equation}
\label{nexp}
n_{ext}(H)=W_v(0)\int_H^{\infty} dS \int da_{ij}  J(a_{ij}) W_{\psi_0,a_{ij}(S,a_{ij})}.
\end{equation}
For the Gaussian field the p.d.f. of the field $\psi_0({\bf x})$
and it's derivative are determined by the correlation function of
$\psi_0({\bf x})$ eq. (\ref{corr}). In equation (\ref{nexp}) we
will integrate over the conditional probability
$W_{con}(a_{ij}/S)=W(a_{ij},S)/W_{\psi_0}(S)$ and will get
\begin{equation}
\label{next}
n_{ext}=W_v(0)\int_H^{\infty} dS W_{\psi_0}(S)\int da_{ij}J(a_{ij})W_{con}(a_{ij}/S).
\end{equation}
Using the properties of Gaussian variables one can receive that the
conditional expected value of $a_{ij}$ is $\langle a_{ij}
\rangle_{con}=S\langle a_{ij} S \rangle/\sigma_{\psi}^2 $. In the
problem of the Burgers turbulence at large time the asymptotic of
$n_{ext}$ of high value $H$ is important. Thus in conditional
averaging we have
\begin{equation}
\langle J(a_{ij}) \rangle_{con}\simeq J(\langle a_{ij} \rangle_{con})\simeq\prod_{i=1}^d
\langle a_{ii}\rangle_{con}=\frac{S^d}{l_{0,eff}^{2d}}.
\label{mass1}
\end{equation}
Here we introduce the effective length $l_{0,eff}$
\begin{equation}
\label{leff}
l_{0,eff}^{d}=\prod_{i=1}^d l_{0,i}
\end{equation}
and take into account that $\langle a_{ii}^2 \rangle=\sigma_{\psi}^2/l_{0,i}^4$.
Finally  we obtain from
equations (\ref{next}),(\ref{mass1}) for the density of extremum the following
expression
\begin{eqnarray}
\label{nextk}
n_{ext}(H)&=&W_v(0)\int^{\infty}_H dS W_{\psi_0}(S)\frac{S^d}{l_{eff}^{2d}}\nn
&=&
\frac{1}{(2\pi)^{\frac{d+1}{2}}l_{eff}^d}\int_{H/\sigma_{\psi}}^{\infty} S^d
e^{-S^2/2\sigma_{\psi}^2}dS\nn
&\simeq&
\left(\frac{H}{\sigma_{\psi}}\right)^{d-1}\frac{1}
{(2\pi)^{\frac{d+1}{2}} l_{eff}^d}e^{-H^2/2\sigma_{\psi}^2}.
\end{eqnarray}
Thus from equation (\ref{nextk}) one can receive that the mean
number of the extremum of anisotropic field $\psi_0(\bf{x})$ is
determined by some effective spatial scale $l_{eff}$, which is
geometrical mean of spatial scale $l_{0,i}$. For the relatively
large $H$ the density of extremum (\ref{nextk}) is equal to the
density of events that the random field $\psi_0({\bf x})$ is over
the $H$.

For the homogeneous field $N(H;V)=Vn_{ext}(H)$, where $n_{ext}(H)$
is described by the equation (\ref{nextk}). For the nonhomogeneous
field, even in one-dimensional case, the expression for $N$ is more
complicated. We assume that the nonhomogeneous field is $\Phi({\bf
x})=\psi_0({\bf x})-\alpha({\bf x})$ and $\alpha({\bf x})$ is a
smooth function in scale of $\psi_0({\bf x})$. Then in a
quasistatic approximation one can receive for the mean number of
events $N(H;V)$ that $\Phi({\bf x})>H$ in a volume $V$ the
following expression
\begin{equation}
\label{last}
N(H;V)=\int_V{n_{ext}(H+\alpha({\bf x}))}dV,
\end{equation}
where $n_{ext}(H)$ is determined by the expression (\ref{nextk})
and is the density of extremum of the statistically homogeneous
function $\psi_0({\bf x})$.

At large time the paraboloid $\alpha= ({\bf x}-{\bf y})^2/2t$ in
equation (\ref{psimax}) is a smooth function in the scale of the
initial potential. Then for the mean number of maximums one can use
quasi-static approximation eq. (\ref {last}) and
\bba
\label{Q-def}
Q(H)&=&\exp{\left(-N_{\infty} (H)\right)}\\
N_{\infty}(H)&=&\int{n_{ext}\left(H+\frac{(\bf{x}-\bf{y})^2}{2t}\right)}dV.
\end{eqnarray}
Here $N_{\infty}(H)$ is the mean number of extremum of $\Phi({\bf
x,y},t)$ in the hole space with magnitude greater then $H$ and
$n_{ext}(H)$ is the density of the number of extremum of the
initial homogeneous potential $\psi_0({\bf x})$ with value greater
then $H$ . For $H \gg \sigma_{\psi}$ the density $n_{ext}(H)$ is
determined by the expression (\ref{nextk}) and we have
\begin{eqnarray}
\label{numb}
N_{\infty}(H)&=&
\frac{1}{(2\pi)^{(d+1)/2}l_{eff}^d}\nn
&&\int{\left(\frac{H}{\sigma_{\psi}}\right)^{d-1}
e^{-(H+{\bf y}^2/2t)^2/2\sigma_{\psi}^2}}d{\bf y}\nonumber\\
&\simeq&\left(\frac{H}{\sigma_{\psi}}\right)^{d-1}
\frac{1}{\sqrt{2}\pi}\left(\frac{\sigma_{\psi} ^2t}{H l_{eff}^2}\right)^{d/2}
e^{-H/2\sigma_{\psi}^2}.
\end{eqnarray}
In equation (\ref{Q-def}) we integrate over the infinite space, but
the effective volume $(\sigma_{\psi}^2t /H l_{eff}^2)^{d/2}$ is
determined by the paraboloid term in equation ( \ref{Q-def}). Now
the effective number of independent local maximum in initial
perturbation is $N_{max}\sim (\sigma_{\psi} t/l_{eff}^2)^{d/2}$ and
increases with time. When $N_{max}\gg 1$ we can introduce the
dimensionless potential $h$ as follows
\begin{equation}
\label{dimlessh}
H=h\sigma_{\psi},\;\;\; h=h_0(1+z/h_0^2),
\end{equation}
where $h_0=H_0\sigma_{\psi}$ and $H_0$ is the solution of the
equation $N(H_0)=1$
\begin{equation}
\label{HOdimoft}
h_0 \simeq d^{1/2}
\left(\log{\frac{\sigma_{\psi} t}{l_{eff}^2(2\pi)^{1/d}}}\right)^{1/2} ,\;\;
\langle \psi({\bf x},t)\rangle\simeq \sigma_{\psi}h_0.
\end{equation}
Thus we have a logarithmic growth of mean potential. The
dimensionless potential has double exponential distribution \bba
\label{doubleexp} Q(z)&=&\exp{(-\exp{(-z)})}\ ,\nn
Q_h(h)&=&\exp{(-\exp{(-(h-h_0)h_0)}}, \eea

One can see that for $t \gg t_{\rm nl}=l_{eff}^2/\sigma_{\psi}$ we
have $N_{max} \gg 1$ and the integral distribution of absolute
maximum is concentrated in narrow region $\triangle
H/H\simeq\sigma_{\psi}^2/H_0^2\ll 1$ near $H_0$. Using this fact one
can get from (\ref{fin-prob}) the probability distribution of the
maximums coordinate
\begin{equation}
\label{probma}
W({\bf y},{\bf x},t)=
\frac{1}{\sqrt{2\pi L^2(t)}}\exp{-\frac{({\bf x}-{\bf y})^2}
{2 L^2(t)}},
\end{equation}
where
\begin{equation}
\label{Length}
L(t)=\left(\frac{\sigma_{\psi} t}{h_0}\right)^{1/2}
=(\sigma_{\psi} t)^{1/2}d^{-1/4}
\left(\log{\frac{\sigma_{\psi} t}{l_{eff}^2(2\pi)^{1/d}}}\right)^{-1/4}
\end{equation}
is the integral scale of the turbulence. From equation
(\ref{vxy}),(\ref{probma}) we see that the one-dimensional
probability distribution of the velocity field is Gaussian and
isotropic. For the energy of each component we have
\begin{equation}
\label{Eiso}
E_i(t)=\sigma_{v,i}^2=\frac{L^2(t)}{t^2}=
\left(\frac{\sigma_{\psi}}{t}\right)d^{-1/2}
\left(\log{\frac{\sigma_{\psi} t}{l_{eff}^2(2\pi)^{1/d}}}\right)^{-1/2}.
\end{equation}
Thus for the anisotropic initial field there's the isotropisation of the
turbulence.

For the multi-dimensional Burgers turbulence the two-dimensional
probability distribution, correlation function and energy spectrum
where found in \cite{GurbatovSaichev1984,GMS91} using so called
``cellular'' model. In this model is assumed that in different
elementary volumes the initial potential are independent and that
the potential has a Gaussian distribution. In this model there's a
free parameter~$\triangle$ which is the size of the elementary
cell. In the present work we consider a continuous initial random
potential field with given correlation function
(\ref{corr}),(\ref{R-ap}). The procedure for calculation of the
two-point probability distribution function is nevertheless
similar to that one used in \cite{GurbatovSaichev1984,GMS91}. It's
easy to show that for the two-point P.D.F. we have the same
expression as obtained in \cite{GMS91} for the cell model, only the
size of the cell $\triangle$ we used to change with the effective
spatial scale $l_{eff}$. The effective spatial scale $l_{0,eff}$
(\ref{leff}) is determined by the scales $l_{0,i}$ of initial
correlation function (\ref{corr}). For the two-point P.D.F.,
correlation function and energy spectrum we also have the
self-similarity and isotropisation at large times. In particular
for the normalized longitudinal and transverse correlation function
of the velocity field $\tilde{\bf v}={\bf v}/\sigma_{v,i}$ we have
\begin{equation}
\tilde B_{LL}(\tilde x)=\langle\tilde v_{1L}\tilde v_{2L}\rangle=
{d\over d\tilde x}\left(\tilde xP(\tilde x)\right),
\label{Blon}
\end{equation}
\begin{equation}
\tilde B_{NN}(\tilde x)=\frac 12\langle\tilde v_{1N}\tilde v_{2N}\rangle=
P(\tilde x),
\label{Bnor}
\end{equation}
where $\tilde x =x/L(t)$ and
\begin{equation}
P(\tilde x) = \frac{1}{2}\int_{-\infty}^\infty {dz \over
g\left({\tilde x+z \over 2}\right) \exp\left[{(\tilde
x+z)^2\over8}\right] + g\left({\tilde x-z \over 2}\right) \exp\left[{(\tilde
x-z)^2\over8}\right]}
\label{defPnoshock}
\end{equation}
\begin{equation}
g(z)\equiv \int_{-\infty}^z e ^{-{s^2\over2}}\,ds.
\label{deferror}
\end{equation}
It may be shown that the function $P(\tilde x)$ is the probability of  having no
shock within an Eulerian interval of length $\tilde x L(t)$.
As far as potential isotropic field are concerned, the normalized energy
spectrum $e(k)$ is formulated via a one-dimensional spectrum $e_{NN}(k)$
of the transverse component. The energy spectrum $E_v(k,t)$ is isotropic and
self-similar
\begin{equation}
E(k,t) = \frac{L^{3}(t)}{t^2} \tilde E(kL(t)).
\label{ssimilarspectrum}
\end{equation}
At large wave number $k$ the discontinuity initiation
leads to the power asymptotic behaviour $E(k,t)\sim k^{-2}$.
At small wave number is also has  the universal behaviour
\begin{equation}
E(k,t) = k^{d+1}\frac{L^{4+d}(t)}{t^2} \sim  k^{d+1} t^{d/2}  ,
\label{ssimsmall}
\end{equation}
which has do with the nonlinear generation of low-frequency component.
In particular for the three-dimensional turbulence we have $E(k,t) \sim  k^{4} t^{3/2}$.
For the large, but finite Reynolds numbers, the ''shocks'' have a finite width
$\delta \sim \nu t /L(t)$ and relative width increases slowly with time
$\delta/L(t) \sim \left(\log{(\sigma_{\psi} t/l_{eff}^2)}\right)^{1/2} $.
Thus at very large time we have a linear stage of evolution.

\subsection{The linear stage of evolution of Burgers turbulence}
\label{s:turblin}

Let us now consider the linear stage of evolution of random field,
when the potential $\psi({\bf x},t)$ and the velocity field ${\bf
v}({\bf x},t)$ eq.(\ref{uhc}) are linearly connected with the
solution $u({\bf x},t)$ of the linear diffusion equation eq.
(\ref{lindif}). Here $u({\bf x},t)$ are the relative fluctuations
of the field $U({\bf x},t)$ eq. (\ref{inu}). Introduce the spectral
density of the field $u({\bf x},t)$ as
\begin{equation}
\label{EuBu} E_u({\bf k},t)=\frac{1}{(2\pi)^d}\int B_u({\bf
z},t)e^{i({\bf kz})}d{\bf z},
\end{equation}
where $B_u({\bf z},t)= \langle u({\bf z},t) u(0,t) \rangle$ is a
correlation function of relative fluctuation field $\tilde U({\bf
x},t)$. The evolution of the spectral density $E_u({\bf k},t)$ and
variance $\sigma_{u}^2(t)$ of $u$ are described by the equations
\begin{equation}
\label{Eukt}
E_u({\bf k},t)=E_{u0}({\bf k}) e^{-2\nu k^2t},\;\;
E_{u0}({\bf k})=E_u({\bf k},0)
\end{equation}
\begin{equation}
\label{sigmau}
\sigma_{u}^2(t)=\int {E_{u0}({\bf k}) e^{-2\nu k^2t}}d{\bf k}\;\;.
\end{equation}
For the homogeneous Gaussian initial potential $\psi_0({\bf x})$ we
have from eq. (\ref{HC})
\begin{equation}
\label{Euk0} E_{u0}({\bf
k})=\frac{1}{(2\pi)^d}\int\left[\exp\left(\frac{B_{\psi}({\bf
z})}{4\nu^2} \right)-1\right]e^{i({\bf kz})}d{\bf z} \;\;,
\end{equation}
where $B_{\psi}({\bf z})$ is the correlation function of initial
potential $\psi_{0}({\bf x})$. The condition of Burgers turbulence
entering the linear regime is $t \gg t_{lin}$, where $t_{lin}$ is
determined from the equation $\sigma_{u}^2(t_{lin}) \simeq 1$ . From
eq. (\ref{Eukt}) one can see, that the old-age behaviour of the
scalar field $u$ and consequently the velocity field ${\bf v}$ will
be determined by the behaviour of the energy spectrum $E_{u0}({\bf
k})$ at small wave numbers ${\bf k}$. When the correlation function
of initial potential $\psi_0({\bf x})$ may be represented in the
form (\ref{corr}) at small ${\bf \rho }$ and
$B_{\psi}(|\rho|>l_{st})\simeq 0$, then from eq. ($\ref{Euk0}$) we
have, that the spectrum  $E_{u0}({\bf k})$ at $k \ll \Re_0/l_{eff}$
is flat and
\begin{equation}
\label{Euk0_0} E_{u0}({\bf k}=0)=D_{u}\simeq
\frac{1}{(2\pi)^{d/2}} (l_{eff}/\Re_0)^d
\exp{\left(\Re_0^2\right)}\;\;,
\end{equation}
where $l_{eff}$ is determined by equation (\ref{leff}). The
spectrum of the field $u$ at large time is isotropic and has an
universal form
\begin{equation}
\label{EuktD}
E_u({\bf k},t)=D_{u} e^{-2\nu k^2t}\;\;\,
\end{equation}
and consequently isotropic is the velocity field
\cite{GurbatovSaichev1984,GMS91,AMS94}. The energy of each
component decays as $E_i(t)\sim D_{u}(\nu t)^{-(d+2)/2}$.

Consider now the case when the initial potential has a correlation
function (\ref{corr}) at small ${\bf x}$ and has a long correlation
at large ${\bf x}$ \cite{AMS94}
\begin{equation}
\label{Bpsilong}
B_{\psi}({\bf x})=
\sigma_{\psi}^2\left(|{\bf x}|/l_{long}\right)^{-\alpha} F_B({\bf x}/|{\bf x}|),\;\;
 \alpha>0.
\end{equation}
Here the function $F_B({\bf x}/|{\bf x}|)$ describes the anisotropy
of correlation function at long distances. If $\alpha>d$ the
evolution of the spectrum of $u$ (eq.\ref{EuktD}) and the velocity
${\bf v}$ will be the same as in absents of long correlation
\cite{AMS94}. At $0<\alpha<d$ the energy spectrum of initial
potential has a singularities at small wave number
\begin{equation}
\label{Epsilong}
E_{\psi}({\bf k})=\sigma_{\psi}^2l_{long}^{\alpha}
|{\bf k}|^{\alpha-d} F_E({\bf k}/|{\bf k}| ),\;\;
 0<\alpha<d,
\end{equation}
where the function $F_E({\bf k}/|{\bf k}|)$ is determined by the
function $F_B({\bf x}/|{\bf x}|)$ and describes the anisotropy of
potential spectrum at small wave number. It was shown \cite{AMS94}
that in this case we have the conservation of anisotropy at linear
stage and asymptotic behaviour of the spectrum of $u$ is determined
by the equation (\ref{Eukt}) where
\begin{equation}
\label{Bulong}
E_{u}({\bf k})=E_{\psi}({\bf k})/(4\nu)^2=\Re_0^2l_{long}^{\alpha}
|{\bf k}|^{\alpha-d} F_E({\bf k}/k)\;\;.
\end{equation}
For the velocity spectrum it means that it reproduced the initial
spectrum of velocity at small wavenumber multiplied by the
exponential factor $\exp{(-2\nu k^2t)}$. These results was
formulated \cite{AMS94} for the correlation function of velocity
fields. In paper \cite{AMS94} also was shown that asymptoticly the
velocity field has Gaussian distribution. But the transformation
processes to the linear stage are not trivial and may be estimated
on base of spectral representation. The behaviour of the spectral
density of the field $u$ at small wave number ${\bf k}$ is
determined by the tail of correlation function $B_{\psi}({\bf x})$
(\ref{Bpsilong}) and from eq. (\ref{Euk0}) we have that the
spectrum is described by the equation (\ref{Bulong}). But with the
increasing of the module of wave number $k$ the power anisotropic
spectrum transformed to the flat spectrum (\ref{Euk0_0}). The wave
number $k_{af}$ where this transformation take place may be
estimated from the condition that at $k \simeq k_{af}$ these
spectrum are the same order and we have
\begin{equation}
\label{kaf}
k_{af}\simeq \Re_0^{(d+2)/(d-\alpha)}
\left(\frac{l_{long}^{\alpha}}{l_{eff}^{d}}\right)^{1/(d-\alpha)}
e^{-\Re_0^2/(d-\alpha)}.
\end{equation}

The condition of Burgers turbulence entering the linear regime is
determined from the equation $\sigma_{u}^2(t_{lin}) \simeq 1$. The
main contribution in the variance $\sigma_{u}^2(t)$ eq.
(\ref{sigmau}) we have from the flat spectrum eq.(\ref{Euk0_0}).
The wave number $k_{lin} \simeq (\nu t_{lin})^{-1/2} $ is
\begin{equation}
\label{klin}
k_{lin}\simeq (\Re_0/l_{eff}) e^{-\Re_0^2/d}.
\end{equation}
Thus for the ratio of this two critical wave numbers we have
\begin{equation}
\label{rkaf} k_{af}/k_{lin} \simeq
\Re_0^{(\alpha+2)/(d-\alpha)}
\left(\frac{l_{long}}{l_{eff}}\right)^{1/(d-\alpha)}
e^{-\Re_0^2\alpha/d(d-\alpha)},
\end{equation}
and for the large initial Reynolds number $k_{af} \ll k_{lin}$. From
the equation $\sigma_{u}^2(t_{lin}) \simeq 1$ we have that $t_{lin}
\simeq t_{\rm nleff}\Re_0^{-1}\exp{(2\Re_0^2)}$ and is extremely
large in compare with the effective nonlinear time $t_{\rm
nleff}\simeq l_{eff}^2/\sigma_{\psi} $. It is easy to see from
(\ref{kaf})) that the time of ``isotropisation'' $t_{iso} \simeq \nu
k_{af}^2$ is much greater then the nonlinear time $t_{iso} \sim
t_{\rm nleff} e^{2\Re_0^2\alpha/d(d-\alpha)}$ and this difference
increaces when the index $\alpha$ is near the critical value $\alpha
\simeq d$.

\section{Discussion and conclusion}

Let us now discuss the evolution of the turbulence in
presence of anisotropy at small (\ref{corr}) or large
(\ref{Bpsilong}) spatial scales. In initial perturbation
the energy of the velocity component is
$\sigma_{\psi}^2/l_{0,i}^2$ and is greater for the small
scales $l_{0,i}$. At the initial stage the scale of the
turbulence in this direction increases faster then in
others and we have primarily the energy decay primarily of
this component (see Section\ref{s:turbin}). After the time
$t$ is greater the of the nonlinear time of the the
component with the largest scale $t_{{\rm nl},i}=\max
l_{0,i}^2/\sigma_{\psi}$ we have the isotropisation of
turbulence in the scales in order of the integral scale of
turbulence $L(t)$ (\ref{Length}). In Section
\ref{s:isotrop} we consider the situation in absence of
long scale correlation. But based on the results of one
dimensional case \cite{GSAFT97} we may suggest that there
are no influence a long scale correlation on the evolution
of the energy. Nevertheless we still have conservation of
anisotropy at large scales $|{\bf x}|\gg L(t)$
(\ref{Bpsilong}). In the spectral representation we have
the conservation of the inital spectrum and anysotropy at
small wave number, but at $k \sim k_s(t) \sim t^{-p}$ the
initial spectrum trasformes into selfsimular spectrum
(\ref{ssimilarspectrum}) with the universal behaviour
$E(k,t) \sim k^{(d+1)}$ eq. (\ref{ssimsmall}) at $kL(t)<1$.
Let us define an energy wavenumber $k_L(t) = L^{-1}(t)\sim
(t\sigma_\psi)^{-1/2}$, which is roughly the wavenumber
around which most of the kinetic energy resides. Hence, the
switching wavenumber $k_s(t)$ goes to zero much faster than
the energy wavenumber. Taking into account the finite
viscosity we have that after the $t\gg t_{lin} \simeq
t_{\rm nleff}\Re_0^{-1}\exp{(2\Re_0^2)}$ the nonlinear evolution
of the spectrum is frozen and only the linear decays of the
small scales is significant (\ref{Eukt}). The frozen
spectrum of the velocity potential has a critical
wavenumber $k_{af}$ bellow which the spectrum is
anisotropic and reproduce the initial spectrum, and at
$k>k_{af}$ is flat eq.(\ref{Euk0_0}) . Thus at $t_{af} \gg
t \gg t_{lin}$ this part of the spectrum will play the
dominate role in the evolution of the velocity spectrum,
consequently the energy of all velocity component are
equal. At $t \gg t_{af}$ the spectrum of the velocity will
be reproduced the small scale part of the initial velocity
spectrum multiplied by the exponential factor (\ref{Eukt}),
and we have finally the anisotropic field.

\vspace{2mm} \par\noindent {\bf Acknowledgments.} We have benefited
from discussion with U.~Frisch, A.~Saichev, A.~Sobolevskii. This
work was supported by Grants\,: RFBR-08-02-00631-a, SS 1055.200.2.
S.~Gurbatov thanks the French Ministry of Education, Research and
Technology for support during his visit to the Observatoire de la
C\^ote d'Azur.

\end{document}